# Modification and the Performance Enhancement of Solar Biomass Dryer

By Tanvir Ahmed & Tanjid Zaman

*University of Engineering and Technology*

*Abstract-* Drying is an important agricultural process, particularly for crops, and shriveled products are used all over the world. The performance of drying green chili was also tested in this article, which created an alternate way of drying agricultural products. The goal of this study is to provide a solar biomass hybrid dryer with improved design, construction, and performance testing. During most hours of the trial, the temperature within the solar collector and dryer was sufficiently higher than the ambient temperature, according to the results obtained during the test period. The temperature of the ambient air at the collector intake ranged from 30 to 35 degrees Celsius. The temperature of the air at the collector's outlet ranged from 54 to 64 degrees Celsius, while the temperature of the drying chamber ranged from 51 to 60 degrees Celsius, making it suitable for drying green chili and a variety of other agricultural products. The collector was found to be 46.54 percent efficient. The findings revealed that the alteration of the collector, which produces turbulent air flow and improves chamber wall insulation, affects drying. Based on the results of this study, the created solar biomass hybrid drier is cost-effective for small-scale crop growers in rural areas of developing countries.

*GJRE-A Classification:* FOR Code: 091399p

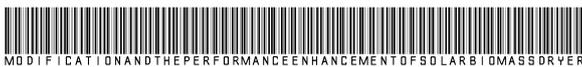

MODIFICATIONANDTHEPERFORMANCEENHANCEMENTOFSOLARBIOMASSDRYER

*Strictly as per the compliance and regulations of:*

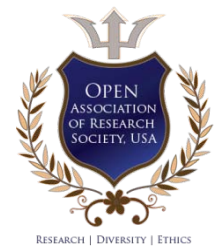

# Modification and the Performance Enhancement of Solar Biomass Dryer

Tanvir Ahmed [α] & Tanjid Zaman [σ]

*Abstract-* Drying is an important agricultural process, particularly for crops, and shriveled products are used all over the world. The performance of drying green chili was also tested in this article, which created an alternate way of drying agricultural products. The goal of this study is to provide a solar biomass hybrid dryer with improved design, construction, and performance testing. During most hours of the trial, the temperature within the solar collector and dryer was sufficiently higher than the ambient temperature, according to the results obtained during the test period. The temperature of the ambient air at the collector intake ranged from 30 to 35 degrees Celsius. The temperature of the air at the collector's outlet ranged from 54 to 64 degrees Celsius, while the temperature of the drying chamber ranged from 51 to 60 degrees Celsius, making it suitable for drying green chili and a variety of other agricultural products. The collector was found to be 46.54 percent efficient. The findings revealed that the alteration of the collector, which produces turbulent air flow and improves chamber wall insulation, affects drying. Based on the results of this study, the created solar biomass hybrid drier is cost-effective for small-scale crop growers in rural areas of developing countries.

## I. Introduction

Vegetables, fruits, and harvests are necessary for long-term storage without compromising product quality. The most common method for drying agricultural products is unsafe open-air sun drying. However, it has a severe problem with wind dust and infestation, and the product may be seriously damaged to the point where it loses market value, resulting in a loss of food quality that could have distressing economic consequences on the local and international market. When compared to open sun drying, solar dryers can reduce the drying time by about 65 percent, improve the quality of the dried product in terms of hygienic, safe moisture content, cleanliness, color, and taste, and protect the product from dust, rain, and insects. The payback period ranges from 2 to 4 years, depending on the rate of utilization.

## II. Literature Review

The features of thin layer solar drying of Brooks and Amelie mangoes were investigated by Dissa [1]. Purohit found a cost-benefit analysis of a solar drier for drying agricultural produce in India [2]. Ferreira studied the technological feasibility of solar drying for agronomical products, namely harvests, in Brazil [3]. Solar drying, according to Tiwari and Barnwal, is an excellent method of food preservation because the product is protected from dust, rain, animals, and insects while drying [4]. Amer has developed and tested a new hybrid sun drier for banana drying [5]. For tomato drying, Hossain has developed a prototype hybrid solar dryer [6]. E. Tarigan created a hybrid drier with natural convection and a biomass backup heater [7]. A direct solar–biomass drier was also created by Vijay and Prasad [8]. Bangladesh has an annual solar radiation availability of up to 1700 kwh/m2. Rajshahi receives the most sun radiation, with measurements ranging from 180.30 to 250 cal/cm2/min. Rajshahi had the lowest cloud broadcasting, with values ranging from 0.34 to 6.36 okta [9]. Also, Various numerical models have been developed to improve the performance of a solar biomass dryer by using the pre-heat method [10, 11]. Improved pyrolysis system can also increase the efficiency of the solar biomass dryer [12].

Author α: Department of Industrial & Production Engineering, Bangladesh University of Engineering and Technology.
e-mail: tanvirahmed12111998@gmail.com
Author σ: Department of Mechanical Engineering, Bangladesh University of Engineering and Technology.
e-mail: sakif.azmain.sa@gmail.com

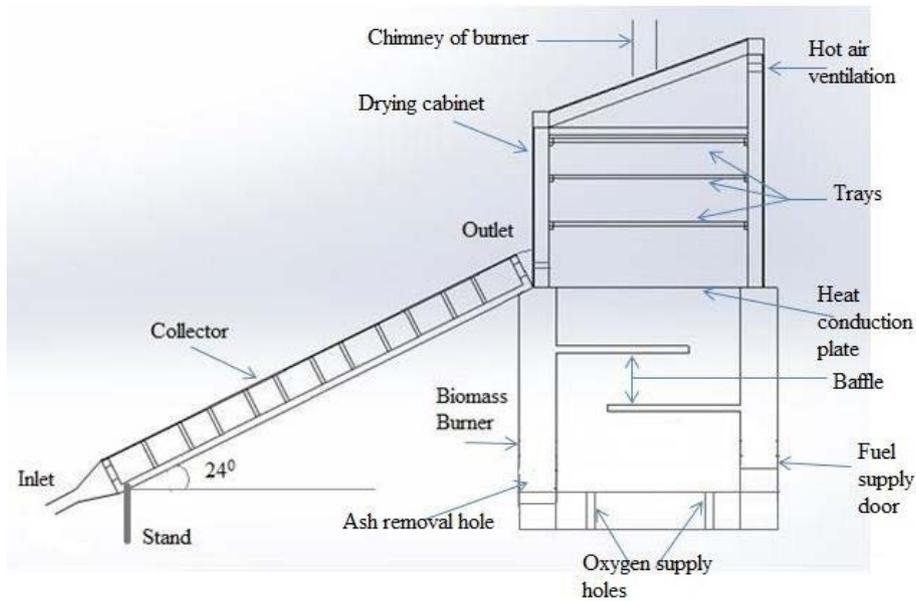

*Fig. 1:* The schematic view of the experimental setup

However, if combined heating from solar and charcoal firings is provided, it will be more efficient [13, 14]. The goal of this study is to create a hybrid solar drier with a biomass backup heater that can achieve the highest heat transfer for drying green chili. Two metal baffle plates were employed within the burner to maximize heat transfer to the burner wall and to prolong the existence of exhaust gasses inside the burner. The burner materials were the bricks that store heat during biomass combustion and give the stored heat to the dryer during inclement weather, such as a cloudy day or at night, resulting in a rapid boost in efficiency.

## III. Design of Studied System

A biomass burner with chimney, a centrifugal blower, a solar air heating collector, and a drying chamber make up the solar dryer. Figures 1 and 2 depict the experimental setup's schematic view and front view, respectively.

The solar collector consists mostly of a top open box with a 5 mm thick transparent glass cover. To reduce heat losses from the collector atmosphere's internal area to the exterior surface, a 0.019 m thick insulation system was introduced. Because Rajshahi's latitude is 24 degrees, the collector's inclination angle is 24 degrees. The drying cabinet and dryer frame were made of well-seasoned woods that could survive insects and the elements. The rectangular box cabinet, which measured 0.76m x 0.63m x 1.17 m, had a back outlet vent to create and control the convection flow of air through the dryer. For added heating, the cabinet's ceiling was covered with 5mm thick transparent glass sheets. The drying trays in the drying chamber were made of a double layer of fine chicken wire mesh with a somewhat wide structure to allow drying air to travel through the food.

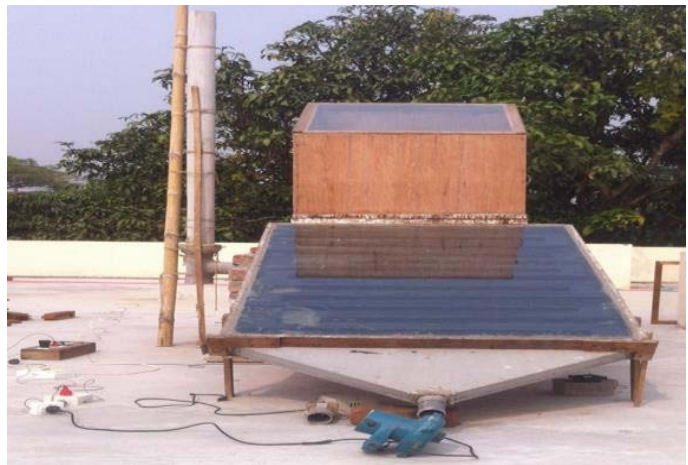

*Fig. 2:* Photograph of the Experimental Setup

## IV. Design Modification of Solar Collector

The design of the solar collector in the present arrangement was changed because it was not adequate. Because the fins inside the collector were arranged in increasing height order, the distance between each fin and the transparent glass cover varied. Because of the huge space between the fin and the glass cover, air travelling through the collector from the blower took less time because it passed over the fins rather than through them. As a result, the air consumed less heat. Because the lower half of the collector was made of plywood, heat was easily dissipated, lowering the efficiency.

*Table 1:* Design parameter of solar collector

| Parameters | Values |
|---|---|
| Area | 1.61 m$^2$ |
| Length | 1.51 m |
| Width | 1.04 m |
| Surface treatment | Black paint |
| Absorber plate | Aluminum sheet |
| Glazing | Normal glass of thickness 5 mm |
| Black insulation | Cork sheet of thickness 25 mm, glass wool |
| Casing | Wood |
| Collector tilt | 24$^0$ |
| Distance between fins and transparent glass | 0.5 inch |
| Fin height | 2.50 inch |

The solar collector's design parameters are listed in Table 1. Some alterations had been made for the sake of improvement.

- Fin height was increased uniformly to 2.50 inches instead of increasing height
- The transparent glass cover was adjusted at a distance 0.5 inch above the fins
- Wood was used for construction of lower part instead of plywood
- The thickness of insulation was increased, and insulation was done more effectively
- Aluminum sheet was used for absorber plate in the place of GI sheet.

## V. Experimental Procedure

In several modes of operation, the experiment was carried out in the forced convection solar biomass hybrid dryer. A thermometer was used to measure ambient temperatures, collector output temperature, and drying chamber outlet temperature every half hour of drying. The beginning and end weights of the goods, as measured by an electronic balance, were the only remaining results noticed. An anemometer was placed between the blower and the collector to monitor the air flow rate. During the experiments, the mass flow rate of air was 0.03 kg/s. The experiment was performed on different solar days in both a natural convection under open sun and a biomass dryer with a forced convection mode, with the results compared to the open sun drying results.

## VI. Instrument used in Experiment

The experiment was carried out with a variety of instruments, which are indicated in table 2.

*Table 2:* Instrumentations used in experiment

| SL No. | Parameters to be measured | Instruments | Accuracy |
|---|---|---|---|
| 1 | Temperature | Thermometer | ± 0.5℃ |
| 2 | Solar radiation | Photovoltaic trainer | ±1 w/m$^2$ |
| 3 | Air velocity | Anemometer | ± 2.5% |
| 4 | Mass | Electronic Balance | 0.01g |
| 5 | Air supply | Blower | N/A |

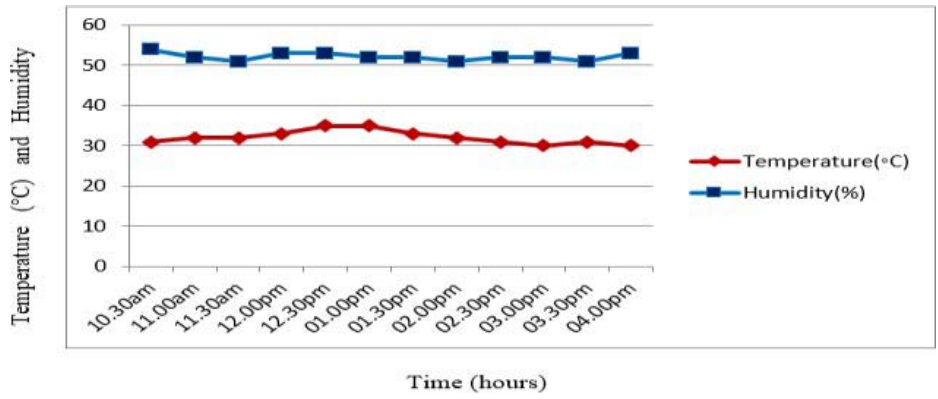

*Figure 3:* Variation of Humidity & Temperature corresponding to the time of the day

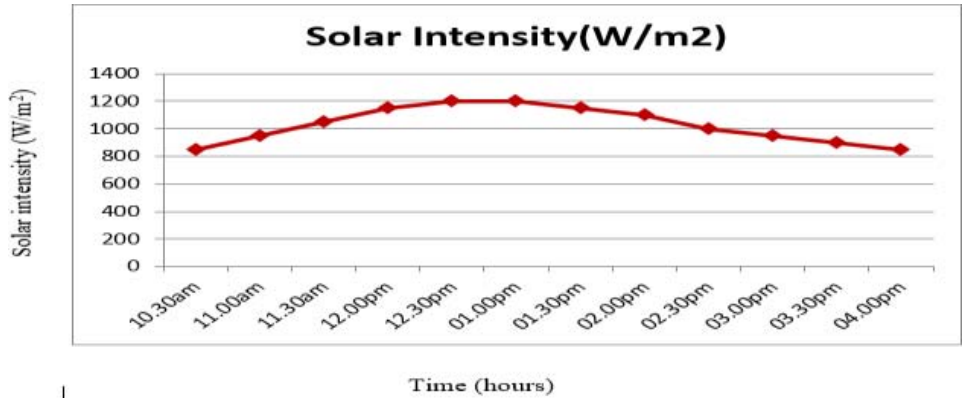

*Figure 4:* Variation of solar intensity corresponding to the time of the day

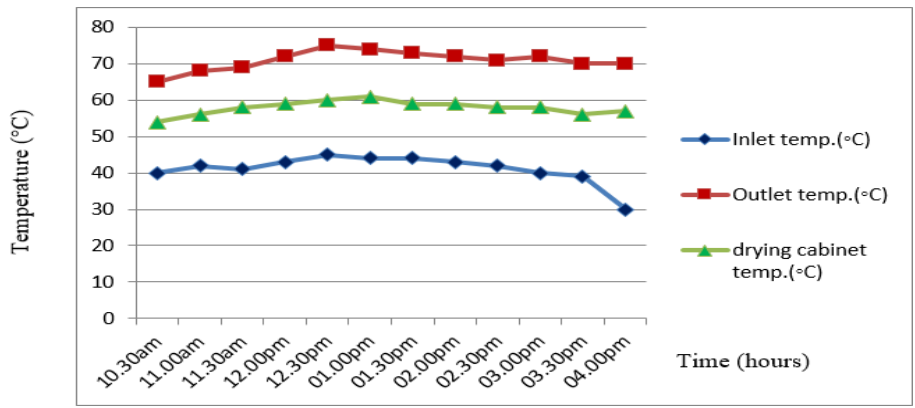

*Figure 5:* Variation of Inlet, outlet and drying chamber temperature recorded in the month June

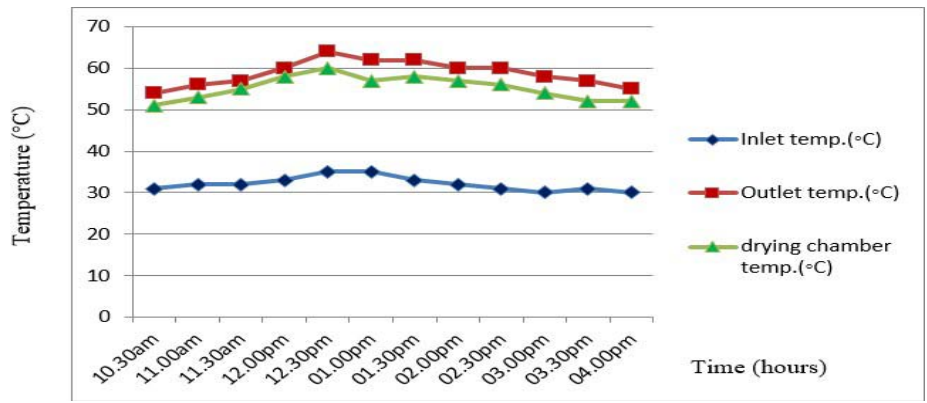

*Figure 6:* Variation of Inlet, outlet and drying chamber temperature recorded in the month October

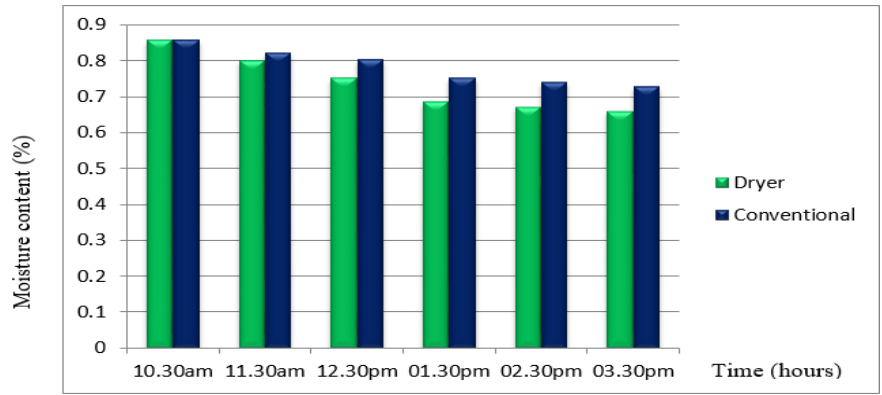

*Figure 7:* Variation of moisture content with time of existing setup

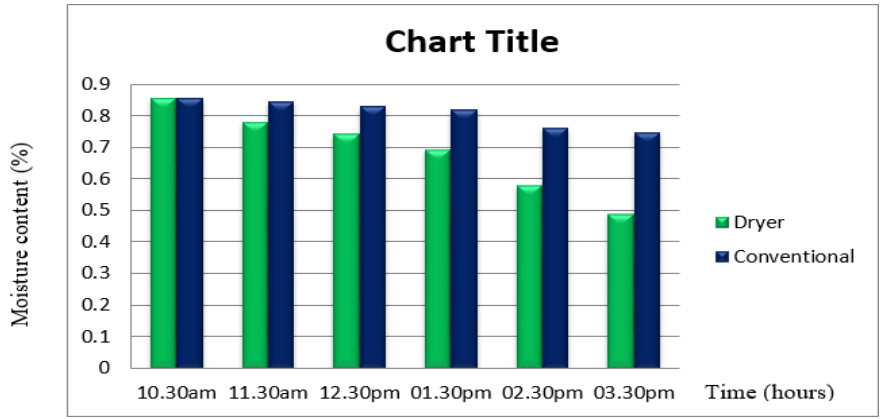

*Figure 8:* Variation of moisture content with time of Modified new

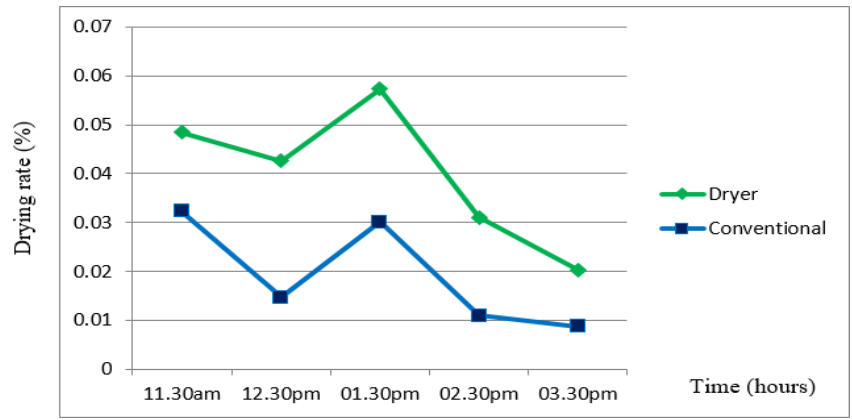

*Figure 9:* Drying rate with time of the day for existing setup (June, 2016)

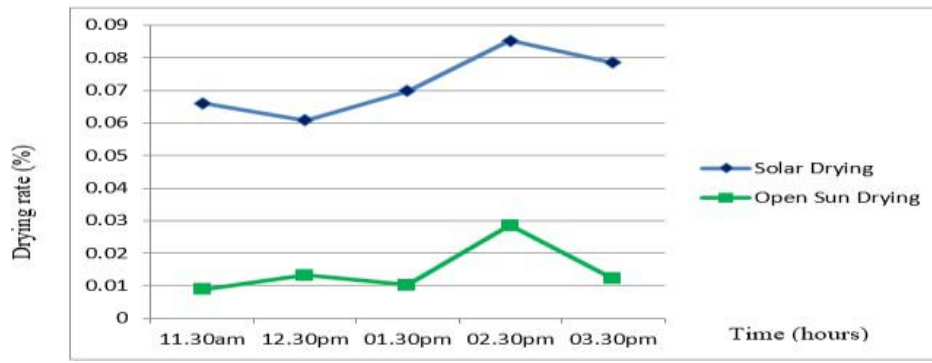

*Figure 10:* Drying rate with time of the day for modified setup (October, 2016)

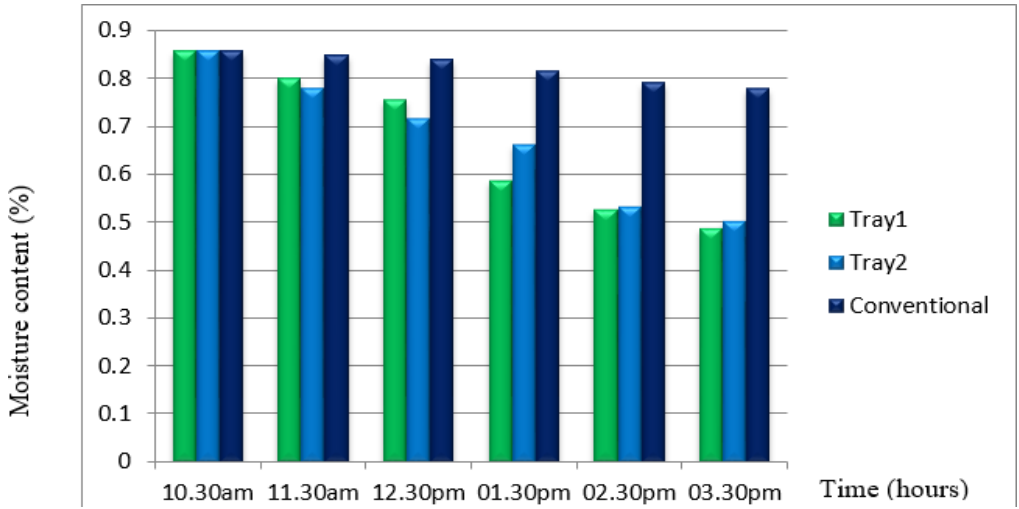

*Figure 11:* Variation of moisture content in different trays and conventional mode with time with Modified new setup

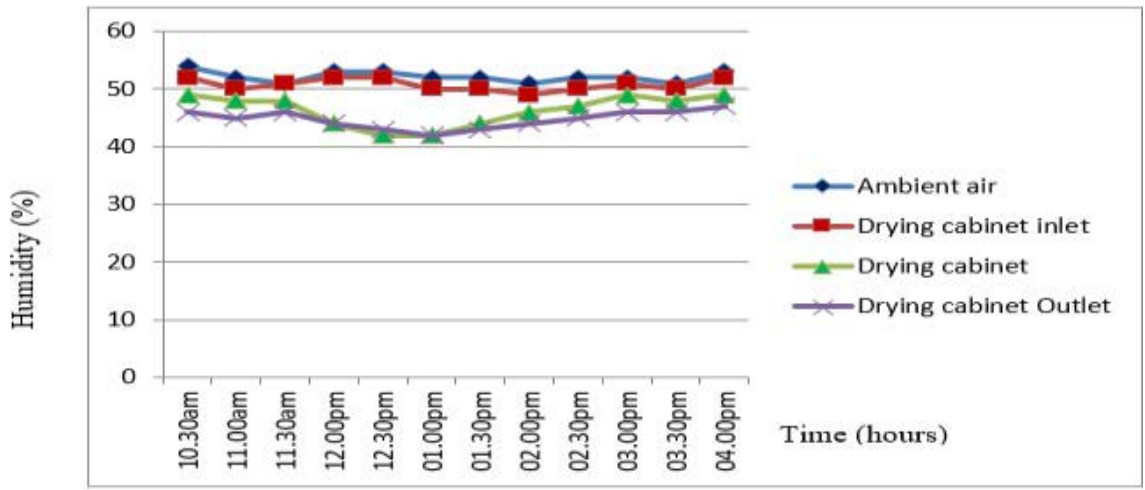

*Figure 12:* Variation of humidity of inlet air of collector, outlet air of collector, ambient air, drying cabinet and drying cabinet outlet

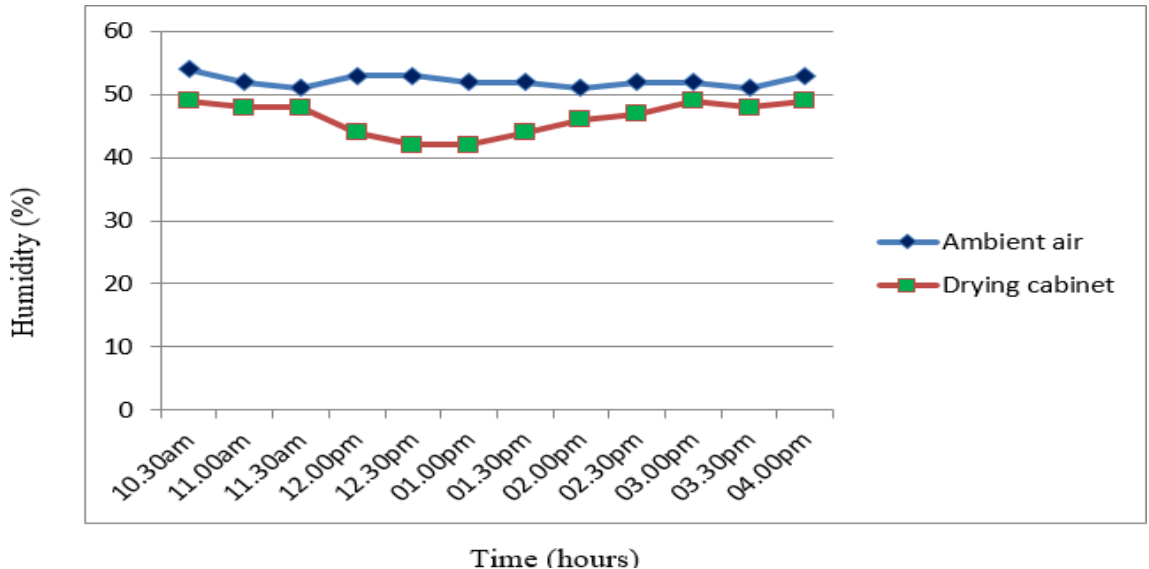

*Figure 13:* Variation of humidity of ambient air and inside of drying cabinet

## VII. Result & Discussion

The experiments were performed in rajshahi, bangladesh (latitude= 24°22´n and longitude= 88°36´e) during 2016. Performance test of the existing solar collector was done during the month june and the performance test of modified collector was done during september and october.

- Weather Condition

Figure 3 depicts the decreasing rate of humidity as the temperature rises. Solar radiation fluctuated from 850 w/m2 to 1200 w/m2 during the experiment, which was monitored using a solar photovoltaic trainer. Figure 4 depicts the daily variation in sun intensity. Maximum sun radiation was recorded during the hours of 10.30 a.m. and 4.00 p.m.

- Performance test of solar collector

Figures 5 and 6 illustrate the results achieved by running the existing solar collector from 10.30 a.m. to 4 p.m. in june and october 2016, respectively, at a mass flow rate of 0.03 kg/s. On the test day, the ambient air temperature at the collector intake ranged from 30 to 45 degrees celsius. The temperature of air at the collector's outlet ranges from 65 to 75° c, while the temperature in the drying chamber ranges from 54 to 61° c. On the test day in october, however, the ambient air temperature at the collector intake ranged from 30o to 35o c. The temperature of air at the collector's outlet ranged from 54 to 64 degrees celsius, while the temperature in the drying chamber ranged from 51 to 60 degrees celsius, which is ideal for drying green chili.

- Comparison with respect to moisture content

Figures 7 and 8 demonstrate the beginning moisture content and ultimate moisture content of the product utilized, green chilli, in both conventional and solar drying modes. The initial moisture content of 1kg of green chilly was 85 percent, and the ultimate moisture content was 65 percent using the previous solar collector. The final moisture content attained after adjustment was 49 percent. It was possible to reduce roughly 16 percent of the extra moisture from green chilly by making these modest changes.

- Comparison with respect to drying rate

Figures 9 and 10 depict the drying rate as a function of time for the original and modified setups, respectively. The average drying rate from the old solar collector was 0.039932 kg/hr, while the average drying rate from the improved solar collector was 0.07211 kg/hr. The weight loss was larger in solar drying mode than in traditional sun drying, according to the findings.

- Performance test of modified solar collector

Figure 11 shows the changes in moisture content in different trays and conventional mode over time in october 2016 with a modified solar collector by placing 1kg of green chilly in the upper tray, referred to as tray1, 1kg in the bottom tray, referred to as tray2, and 1kg in the open atmosphere for conventional drying mode. Each person's weight was collected at one-hour intervals until 3.30 p.m. it demonstrates that tray 1 has a lower moisture removal rate than tray 2 until 12.30 pm, after which tray 1's moisture removal rate has increased. Because tray1 is exposed to direct sunlight as well as heat from the solar collector, and because the intensity of the sun is highest between 12.30 pm and 2.00 pm, the moisture removal rate of tray1 is increasing at that time.

- Variation of humidity

In the month of october, figure 12 depicts the fluctuation in humidity of the collector's inlet air, outlet air, ambient air, drying cabinet, and drying cabinet outlet in (percent) as a function of time of day. The graph indicates that as the temperature rises, the humidity falls. During the day, the highest humidity of ambient air is 54 percent and the minimum is 50 percent, as shown in the graph. The humidity in the drying cabinet inlet ranges from 49 to 52 percent. The humidity of ambient air and the humidity in the drying cabinet are compared in figure 13.

## VIII. Conclusion

The solar collector of a forced convection type solar dryer with a backup biomass heater was updated in design for improved efficiency and used for drying. The drying performance of green chili was examined using the original setup and the modified new setup in three different modes of operation: conventional drying, sun drying, and hybrid drying. Depending on the mode of operation, the temperature within the drying chamber ranges from 44 to 75° c. The average efficiency of an existing sun collector was 37.49 percent, with outlet air temperatures ranging from 52 to 71 degrees celsius, but with the improved new solar collector, the efficiency was 46.54 percent. The entire reduction in traditional sun mode drying was 12.94% in the present setup, whereas the total reduction in solar drying was 36.6%. The total reduction in traditional sun mode drying was 10.97%, whereas the total reduction in solar drying was 20.79%. As a result, the change of the solar collector was beneficial to the total setup, as the collector's efficiency increased from 37.49 percent to 46.54 percent. The performance was satisfactory, and following the modifications, we were able to overcome the issue of not having enough sun radiation during the day by obtaining consistent drying, which is desirable for most agricultural products.

## References Références Referencias